\documentclass[epj]{webofc}
\usepackage[varg]{txfonts}   % Web of Conferences font

\usepackage{bm}
\usepackage{amssymb}
\usepackage{amsmath}
\usepackage{enumerate}
\usepackage{epsf}
\usepackage{epstopdf} %%%%%%%%%%%%%%%%%%%%

\def\eRM{{\mathrm e}}
\def\dRM{{\mathrm d}}

\def\mx{{\bm x}}

\def\eps{\varepsilon}

\allowdisplaybreaks
\allowbreak
\newcommand{\fp}[1]{FP$ {\textrm{#1}}$}

\begin{document}

\selectlanguage{english}
\title{Stochastic Navier-Stokes equation and advection of a tracer field: One-loop renormalization near $d=4$}
% subtitle is optional

\author{N. V. Antonov\inst{1}\fnsep\thanks{\email{n.antonov@spbu.ru}} \and
        N. M. Gulitskiy\inst{1}\fnsep\thanks{\email{n.gulitskiy@spbu.ru}} \and
        M. M. Kostenko\inst{1}\fnsep\thanks{\email{m.m.kostenko@mail.ru}} \and
        T. Lu\v{c}ivjansk\'y\inst{2}\fnsep\thanks{\email{tomas.lucivjansky@uni-due.de}}
        }

\institute{Department of Theoretical Physics, Faculty of Physics,  
Saint-Petersburg State University, 7/9~Universitetskaya nab., St. Petersburg, 199034 Russia
\and
Faculty of Sciences, P.J. \v{S}af\'arik
University, Moyzesova 16, 040 01 Ko\v{s}ice, Slovakia
}

\abstract{
The renormalization group approach and the operator product expansion technique are applied to the model of a tracer field advected by the Navier-Stokes velocity ensemble for a compressible fluid. The model is considered in the vicinity of the specific space dimension $d=4$. The properties of the equal-time structure functions are investigated. The multifractal behaviour of various correlation functions is established. All calculations are performed in the leading one-loop approximation.

}

\maketitle
%%%%%%%%%%%%%%%%%%%%%%%%%%%%%%%%%%%%%%%%%%%%

\section{Introduction}

The investigation of such behavior as anomalous scaling requires a lot of thorough, even meticulous, analysis to be carried out. The phenomenon manifests itself in a singular (arguably, power-like) behaviour of some statistical quantities (correlation functions, structure functions, etc.) in the inertial-convective range in the fully developed turbulence regime~\cite{Legacy}.

Very useful and computationally efective approach to the problems with many interacting degrees of freedom on different scales is the field-theoretic renormalization group (RG) approach which can be subsequently accompanied by the operator product expansion (OPE); see the monographs~\cite{Vasiliev,Zinn,turbo,Tauber}. 
One of the greatest challenges is the investigation of the Navier-Stokes equation for a compressible fluid, and, in particular, 
passive scalar field advection by this velocity ensemble. The first relevant discussion and analyses of the problem of passive advection emerged a few decades ago for the Kraichnan's velocity ensemble~\cite{Kraich1,GK,RG}. Further studies are connected with its more realistic generalizations~\cite{JphysA,JJ15,V96-mod,amodel,AG12,J13,Uni3,VectorN,HZ,Marian16}; see also review paper~\cite{HHL}. The RG+OPE technique were also applied to more complicated models, in particular, to the compressible case~\cite{NSpass,Ant04,AK14,AK15,AGM,JJR16,VM,tracer2,tracer3,AK2,LM,St,MTY,ANU97}.

The present study is devoted to the investigation of the passive scalar tracer field advected by the Navier-Stokes velocity ensemble with the compressibility taken into account near the specific space dimension~$d=4$. Usually $d$ plays a passive role~-- see~\cite{ANU97} for Navie-Stokes equation itself and~\cite{AK14,AK15} for advection of scalar fields~-- but if $d=4$, the situation is similar to the analysis of the incompressible Navier-Stokes equation near space dimension $d=2$. In this case an  additional divergence appears in the 1-irreducible Green's function $\left\langle v'v'\right\rangle_{\text{1-ir}}$, see~\cite{HN96,AHKV03,AHH10}. This feature allows us to employ a double expansion scheme, in which the formal expansion 
parameters are $y$, which describes the scaling behaviour of a random force, and $\eps=4-d$, i.e., a deviation from the space dimension $d=4$.

The paper is a logical continuation of our previous studies~\cite{AGKL,AGKL2} and is organized as follows. In the introductory Section~\ref{sec:fp} a brief discussion of the fixed points' structure and possible scaling regimes is presented. In the following section the tracer field is introduced and the expressions for its critical dimensions (for each of two scaling regimes) are presented. In Section~\ref{sec:compf} the renormalization of a certain composite field is considered. In Section~\ref{sec:ope} OPE is applied to the equal-time structure functions constructed of the tracer fields; the anomalous exponents are calculated. The concluding section is devoted to the brief discussion.

\section{Navier-Stokes equation, scaling regimes}\label{sec:fp}

Consider first an advection of a scalar field by a compressible fluid. The fluid within the model used is assumed to be compressible, so the Navier-Stokes equation and the continuity equation have the forms
\begin{align}
\nabla_{t} v_{i} &=
\nu_{0} [\delta_{ik}\partial^{2}-\partial_{i}\partial_{k}]
v_{k}\! +\! \mu_0 \partial_{i}\partial_{k} v_{k} -\!
\partial_{i} \phi\! +\! f_{i},  \quad \nabla_t \equiv \partial_t + v_i\partial_i,
\label{ANU} \\
\nabla_{t} \phi &= -c_{0}^{2} \partial_{i}v_{i}.
\label{ANU1}
\end{align}
Here $\partial _t \equiv \partial /\partial t$,
$\partial _i \equiv \partial /\partial x_{i}$, $v_i=v_i(t,\mx)$ is a velocity field, 
 $\phi = \phi(t,\mx)$ is a pressure field and $\partial^{2}=\partial _i\partial _i$
is the Laplace operator. The constants $\nu_{0}$ and $\mu_{0}$ are two
independent molecular viscosity coefficients, $c_0$  is an adiabatic speed of sound.
The statistics of the random force $f$ isconveniently given in the Fourier
representation  by the relation
\begin{equation}
  \langle f_i(t,{\bm x}) f_j(t',{\bm x}') = \frac{\delta(t-t')}{(2\pi)^d} \int_{k>m} \dRM^d{\bm k}\mbox{ } {D}_{ij}({\bm k})
  \eRM^{i{\bm k}\cdot({\bm x-\bm x'})},
\label{ff}
\end{equation}
where the delta function ensures Galilean invariance of the model and $m$
 provides us with infrared~(IR) cutoff. Further, the kernel function ${D}_{ij}({\bm k})$ reads
\begin{equation}  
  D_{ij}({\bm k})=g_{10} \nu_0^3 k^{\varepsilon-y} \biggl\{
  P_{ij}({\bm k}) + \alpha Q_{ij}({\bm k})
  \biggl\} 
  + g_{20} \nu_0^3 \delta_{ij}.
  \label{eq:correl2}
\end{equation}
A term proportional to the charge $g_{10}$ ensures a steady input of energy into the system, which
is needed in order to counterbalance dissipation processes due to a viscosity; the definition $\varepsilon=4-d$ is introduced here. The term proportional to $g_{20}$ is not dictated by the physics of the fluid, but rather by a proper renormalization treatment~\cite{BigOne}.

In \cite{AGKL} three IR attractive fixed points, which defines possible scaling regimes of the system, were discussed. The character of the IR behaviour depends on the mutual relation between $y$ and $\varepsilon$~-- two formally small quantities which were introduced in the correlator of the random force in the Navier-Stokes equation. In practical calculations
they constitute parameters into which universal quantities are expanded. This is done in a similar
fashion as calculation of critical exponents in $\phi^4$ theory, see~\cite{Vasiliev,Zinn,turbo,Tauber}.

The fixed point \fp{I} (the trivial or Gaussian point) is stable if $y$, $\varepsilon<0$. The coordinates are
\begin{equation}
  g_1^* = 0, \quad g_2^* = 0.
  \label{fp1}
\end{equation}
The fixed point \fp{II}, which is stable if $\varepsilon>0$ and $y<3\varepsilon/2$, has the following coordinates:
\begin{equation}
  g_1^* = 0, \quad g_2^* = \frac{8\eps}{3}.
  \label{fp2}
\end{equation}
The fixed point \fp{III} (stable if $y>0$ and $\varepsilon<2y/3$) is
\begin{equation}
  g_1^* = \frac{16y(2y-3\eps)}{9[y(2+\alpha)-3\eps]}, 
  \quad g_2^* =\frac{16\alpha y^2}{9[y(2+\alpha)-3\eps]}.
  \label{fp3}
\end{equation}
The crossover between the two nontrivial points~(\ref{fp2}) and~(\ref{fp3}) takes place across the line $y=3\eps/2$, which is in accordance with results of~\cite{Ant04}.

Depending on the values of $y$ and  $\varepsilon$, the different values of the critical dimension for various quantities $F$ are obtained. They can be calculated via the expression
\begin{equation}
\Delta[F]=d^{k}_{F}+ \Delta_{\omega}d^{\omega}_{F} + \gamma_{F}^{*},
\label{Delta}
\end{equation}
where $d^{\omega}_{F}$ is the canonical frequency dimension, $d^{k}_{F}$ is the momentum dimension, $\gamma_F^*$ is the anomalous dimension at the critical point (FPII or FPIII), and $\Delta_{\omega}=2-\gamma_\nu^*$ is the critical dimension of frequency.

In \cite{AGKL} the critical dimension of the passive scalar density field $\theta$ and the field $\theta'$ were obtained for the fixed points \fp{II} and \fp{III}:
\begin{equation}
\Delta_{\theta}= -1+\varepsilon/4, \quad \Delta_{\theta'}= d+1 -\varepsilon/4 \quad \text{for the fixed point \fp{II}};
\label{KriTet1}
\end{equation}
\begin{equation}
\Delta_{\theta}= -1+y/6, \quad \Delta_{\theta'}= d+1 -y/6  \quad \text{for the fixed point \fp{III}}.
\label{KriTet2}
\end{equation}

\section{Tracer field}\label{sec:tracer}

There are two permissible kinds of passive scalar fields in nature: the density field (density of some pollutant) and the tracer field which describes the temperature or entropy~\cite{LL}. The advection of the density field is described by the stochastic equation
\begin{equation}
\partial _t\theta+ \partial_{i}(v_{i}\theta)=\kappa _0 \partial^{2} \theta+f,
\label{density1}
\end{equation}
while the advection of the tracer field is governed by
\begin{equation}
\partial _t\theta+ (v_{i}\partial_{i})\theta=\kappa _0\partial^{2} \theta+f.
\label{tracer1}
\end{equation}
Here $\kappa_0$ is the molecular
diffusivity coefficient and $f\equiv f(x)$ is a Gaussian noise with zero mean and given covariance,
\begin{equation}
\langle f(x)f(x') \rangle = \delta(t-t')\, C({\bf r}/L), \quad
{\bf r}= {\bf x} - {\bf x}',
\label{noise}
\end{equation}
where $C({\bf r}/L)$ is some function finite at $({\bf r}/L)\to 0$ and rapidly decaying for $({\bf r}/L)\to\infty$.
For the case of the incompressible fluid the difference between the density and tracer fields is indistinguishable since an additional condition for the velocity field arises (namely, transversality condition $\partial_i v_i = 0$), which makes the terms $\partial_{i}(v_{i}\theta)$ and $(v_{i}\partial_{i})\theta$ identical.

The aim of this study is to investigate the critical behaviour of various statistical quantities which are constructed in terms of the tracer field advected by the compressible fluid near the special space dimension $d=4$. The IR (large-scale) behavior of the structure functions is the major focus of this paper. Within the method used the composite field theory and OPE are employed. This approach requires a strong RG analysis aimed at finding (or discovering) the critical dimension of the tracer field.

Fortunately, despite the difference between the formalism for density and tracer fields in the case of a compressible fluid, 
the critical dimensions for these two fields have the same form; for detailed analysis see~\cite{AK14}. Thus, expressions~(\ref{KriTet1}) and~(\ref{KriTet2}) remain true for tracer field as well. In what follows the composite fields and OPE will be discussed.

\section{Composite fields}\label{sec:compf}

The measurable quantities are some correlation functions or structure functions of composite operators.
A local composite operator is a monomial or polynomial constructed from the primary fields
$\theta(x)$ and their finite-order derivatives at a single space-time point $x$. In the Green's functions with such
objects, new UV divergences arise due to the coincidence of
the field arguments. They can be removed by the additional
renormalization procedure. 

The simplest case of a composite operator is the scalar operator $F(x)=\theta^n(x)$. But in contrast to the density model, if the field $\theta(x)$ obeys stochastic equation~\eqref{tracer1}, the real index of divergency for the diagrams with such operators is positive. This means, that in case of tracer such operators do not diverge and renormalization constants for them are trivial.

More interesting objects are composite operators with derivatives, namely 
\begin{equation}
F^{(n,l)}_{i_{1}\dots i_{l}} =
\partial_{i_{1}}\theta\cdots\partial_{i_{l}}\theta\,
(\partial_{i}\theta\partial_{i}\theta)^{s} + \dots
\label{Fnp}
\end{equation}
Here $l$ is the number of the free vector indices (the rank of the tensor) and $n=l+2s$ is the total number of the fields $\theta$ entering the
operator. The ellipsis stands for the subtractions with the Kronecker's delta symbols that make the operator irreducible (so that a contraction with respect to any pair of the free tensor indices vanish). For instance,
\begin{equation}
F^{(2,2)}_{ij} = \partial_{i}\theta \partial_{j}\theta -
\frac{\delta_{ij}}{d}\, (\partial_{k}\theta\partial_{k}\theta).
\label{Irr}
\end{equation}

For practical calculations, it is convenient to contract the tensors
(\ref{Fnp}) with an arbitrary constant vector
{\mbox{\boldmath $\lambda$}}$=\{\lambda_{i}\}$.
The resulting scalar operator takes the following form:
\begin{equation}
F^{(n,l)} = (\lambda_{i}w_{i})^{l} (w_{i}w_{i})^{s} + \dots,
\quad w_{i} \equiv \partial_{i}\theta,
\label{FnpSk}
\end{equation}
where the subtractions, denoted by the ellipsis, necessarily include the
factors of $\lambda^{2}=\lambda_{i}\lambda_{i}$. 

In order to calculate the critical dimension of the operator, one has to renormalize it. The operators~(\ref{Fnp}) can be treated as multiplicatively renormalizable, $F^{(n,l)} = Z_{(n,l)} F^{(n,l)}_{R}$, with certain renormalization constants $Z_{(n,l)}$ (see \cite{AK14}). The counterterm to $F^{(n,l)}$ must have the same rank as the operator itself. It means that the terms containing $\lambda^{2}$  should be excluded since the contracted fields $w_{i}w_{i}$, standing near them, reduce the number of free indices. It is sufficient to retain only the principal monomial, explicitly shown in~(\ref{FnpSk}),
and to discard in the result all the terms with factors of $\lambda^{2}$.
The renormalization constants $Z_{(n,l)}$ are determined by the finiteness of the 1-irreducible Green's function $\Gamma_{nl}(x; \theta)$, which in the one-loop approximation is diagrammatically represented as follows:
\begin{equation}
\Gamma_{nl}(x; \theta) = F^{(n,l)} + \frac{1}{2}   \vcenter{\hbox{\includegraphics[width=0.07\textwidth,clip]{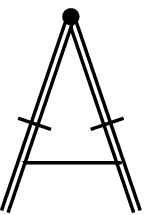}}} ,
\label{dom}
\end{equation}
where numerical factor $1/2$ is a symmetry factor of the graph and the thick dot with two lines attached denotes the operator vertex
\begin{equation}
V(x;x_{1},x_{2})=  \frac{\delta^{2}F^{(n,l)}}{\delta(\partial_{i}\theta)\,\,\delta(\partial_{j}\theta)}.
\label{Vae1}
\end{equation}
The expressions for the propagators and vertices at the bottom of the diagram can be found in~\cite{AGKL,BigOne}. 
Then using the chain rule and up to irrelevant terms the vertex~(\ref{Vae1}) for the operator $F^{(n,l)}$ can be presented in the form 
\begin{equation}
V(x;x_{1},x_{2})=  \frac{\partial^{2}F^{(n,l)}}{\partial w_{i}\partial w_{j}}
\,\partial_{i}\delta(x-x_{1})\, \partial_{j}\delta(x-x_{2}).
\label{Vae10}
\end{equation}
The differentiation yields
\begin{eqnarray}
{\partial^{2}F^{(n,l)}}/{\partial w_{i}\partial w_{j}} &=&
2s (w^{2})^{s-2} (\lambda w)^{l} \left[\delta_{ij} w^{2} +2(s-1)w_{i}w_{j}
\right] + l(l-1) (w^{2})^{s}(\lambda w)^{l-2} \lambda_{i} \lambda_{j}+
\nonumber \\
&+& 2ls (w^{2})^{s-1}(\lambda w)^{l-1} (w_{i}\lambda_{j}+ w_{j}\lambda_{i}),
\label{Vae11}
\end{eqnarray}
where $w^{2}=w_{k}w_{k}$ and $(\lambda w)=\lambda_{k}w_{k}$. Two more
factors $w_{p}w_{r}$ are attached to the bottom of the diagram due to the
derivatives coming from the vertices $\theta'(v\partial)\theta$. The ultraviolet
divergence is logarithmic and one can set all the external frequencies and
momenta equal to zero; then the core of the diagram takes the form
\begin{equation}
\int\frac{\text{d}\omega}{2\pi}
\int_{k>m}\frac{{\text{d}^d}{\bf k}}{(2\pi)^{d}}\,
k_{i}k_{j}\, D_{pr} (\omega, {\bf k})\,
\frac{1}{\omega^{2}+w^{2}\nu^2 k^{4}}.
\label{triad}
\end{equation}
Here the first factor comes from the derivatives in~(\ref{Vae1}), $w=\kappa/\nu$, $D_{pr}$ is the velocity correlation function [see~(\ref{eq:correl2})], and
the last factor comes from the two propagators $\langle\theta'\theta\rangle_{0}$.

After the integration, combining all the
factors, contracting the tensor indices and expressing the result in terms of $n=l+2s$ and $l$, one obtains: 
\begin{align}
\Gamma_{n}(x;\theta) & = F^{(n,l)}(x)\, \biggl\{ 1 - \frac{1}{4wd(d+2)}\,
\biggl[
\frac{Q_{1}}{(1+w)} \biggl(g_{10}\frac{m^{-y}}{y}+g_{20}\frac{m^{-\eps}}{\eps}\biggl) \nonumber\\
& + 
\frac{Q_{2}}{u(u+w)}\biggl(\alpha g_{10}\frac{m^{-y}}{y}+g_{20}\frac{m^{-\eps}}{\eps}\biggl)
\biggl] \biggl\}.
\label{QQ}
\end{align}
Here the following abbreviations have been used:
\begin{eqnarray}
Q_{1} = -n(n+d)(d-1) + (d+1) l(l+d-2), \qquad
Q_{2} = -n(3n+d-4) + l(l+d-2).
\label{Q}
\end{eqnarray}
Then the renormalization constants $Z_{(n,l)}$ calculated in the MS scheme read
\begin{equation}
Z_{(n,l)} = 1 - \frac{1}{2d(d+2)}\,
\left[
\frac{Q_{1}}{2w(1+w)} \left(\frac{g_{10}}{y}+\frac{g_{20}}{\eps}\right) + \frac{Q_{2}}{2wu(u+w)}\left(\alpha \frac{g_{10}}{y}+\frac{g_{20}}{\eps}\right)
\right].
\label{ZZ}
\end{equation}
For the corresponding anomalous dimension one obtains
\begin{equation}
\gamma_{(n,l)} = \frac{1}{2d(d+2)}\, \left\{
\frac{Q_{1}}{2w(1+w)} (g_{10}+g_{20}) +  \frac{Q_{2}}{2wu(u+w)} (\alpha g_{10}+g_{20})
\right\}.
\label{GG}
\end{equation}

In order to evaluate the critical dimension, one needs to substitute the coordinates of the fixed points~\eqref{fp1}~-- \eqref{fp3} into the  expression~(\ref{GG}) and then use the relation~(\ref{Delta}). For the fixed point FPII the critical dimension is
\begin{equation}
\Delta_{(n,l)} = \frac{n}{4}\varepsilon + \frac{Q_1+Q_2}{3d(d+2)}\varepsilon.
\label{Dnl1}
\end{equation}
For the fixed point FPIII it is
\begin{equation}
\Delta_{(n,l)} = \frac{n}{6}y + \frac{2y}{d(d+2)}\frac{Q_1 (\alpha y + 2y - 3\varepsilon) + 3\alpha Q_2 (y-\varepsilon)}{9[y(2+\alpha)-3\varepsilon]}.
\label{Dnl2}
\end{equation}
Both expressions~\eqref{Dnl1} and ~\eqref{Dnl2} suppose higher order corrections in $y$ and $\varepsilon$. 

Therefore, the infinite set of  operators with negative critical dimensions, whose spectra is unbounded from below, is observed.

\section{Operator Product Expansion}\label{sec:ope}

Consider the equal-time structure functions 
\begin{equation}
S_{n}(r) = \langle [\theta(t,{\bf x})-\theta(t,{\bf x'}]^{2n}\rangle
= (\nu\mu^{2})^{-n} \eta(\mu r, mr),
\quad \text{where}\quad r=|{\bf x'}={\bf x}|.
\label{struc}
\end{equation}
The second equality involving the dimensionless functions $\eta$ follows from dimensionality considerations. Solving the RG equations gives the
asymptotic expressions in the limit $\mu r\gg 1$:
\begin{equation}
S_{n}(r) =  (\nu\mu^{2})^{-n} (\mu r)^{-2n\Delta_{\theta}}
\zeta(mr),
\label{struc2}
\end{equation}
where $\zeta$ is some unknown scaling function. The inertial-convective range corresponds to the additional condition $mr\ll1$. The behavior of the
functions $\zeta$ at $mr\to0$ can be studied by means of the OPE.

The tracer field is determined by the stochastic equation~(\ref{tracer1}), therefore, the model is invariant with respect to the constant shift $\theta(x)\to\theta(x)+{\rm const}$. This means, that the operators entering the corresponding OPE,
\begin{equation}
[\theta(t,{\mx})-\theta(t,{\mx'}]^{2n} \simeq \sum_{F} C_{F}(mr)\,
F \biggl(t,\frac{\mx + \mx'}{2} \biggl),
\label{OPE2}
\end{equation}
must be also invariant. Therefore, that they can involve the field $\theta$ only
in the form of derivatives. Clearly, the leading term of the behaviour at small $mr$ should
be determined by the scalar operator with maximal possible
number of the fields $\theta$ (namely, $2n$ for a given $S_{n}$) and the
minimal possible number of spatial derivatives (namely, $2n$~-- one derivative for each $\theta$). Consequently, this is the operator
$F^{(2n,0)} = (\partial_{i}\theta\partial_{i}\theta)^{n}$; see the definition~(\ref{Fnp}).
Thus, the leading-order expression for $S_{n}$ in the inertial range is
\begin{equation}
S_{n}(r) \sim  (\nu\mu^{2})^{-n} (\mu r)^{-2n\Delta_{\theta}}
(mr)^{\Delta_{(2n,0)}},
\label{fin2}
\end{equation}
with the dimensions $\Delta_{(2n,0)}$ given by the expressions~(\ref{Dnl1}) and~(\ref{Dnl2}) with the substitution $l=0$. Other types of fields or other numbers of derivatives generate the more ``distant'' corrections. Expression~\eqref{fin2} together with expressions~\eqref{KriTet1}, \eqref{KriTet2} and~\eqref{Dnl1}, \eqref{Dnl2} means that existence of anomalous scaling, i.e., singular power-like dependence on the integral scale $L$, was established.

Considering OPE for the correlation functions $\langle F^{(p,0)} F^{(k,0)} \rangle$ with $n=p+k$, where $F^{(n,l)}$ is the operator of the type~\eqref{Fnp}, one can observe that the leading contribution to the expansion is determined by the operator $F^{(n,0)}$ from the same family. Therefore, in the inertial range these correlation functions acquire the form
\begin{equation}
\langle F^{(p,0)}(t,{\bf x}) F^{(k,0)}(t,{\bf x'}) \rangle \sim
r^{-\Delta_{(p,0)}-\Delta_{(k,0)}+\Delta_{(n,0)}}.
\label{MF}
\end{equation}
The inequality $\Delta_{(n,0)}<\Delta_{(p,0)}+\Delta_{(k,0)}$, which follows from both explicit one-loop expressions~(\ref{Dnl1}) and~(\ref{Dnl2}), indicates, that the operators $F^{(n,0)}$ demonstrate a ``multifractal'' behavior; see \cite{DL}.

\section{Conclusion}\label{sec:conclusion}

In the present paper the advection of the passive scalar tracer field by the Navier-Stokes velocity ensemble has been examined. The fluid was assumed to be compressible and the space dimension was close to $d=4$. The problem has been investigated by means of renormalization group and operator product expansion; the double expansion in $y$ [see~\eqref{eq:correl2}] and $\varepsilon=4-d$ was constructed. The present study has been aimed at the investigation of the anomalous scaling in the equal-time structure functions for the tracer field. 

There are two IR stable fixed points in this model and, therefore, the critical behaviour in the inertial range demonstrates two different regimes depending on the relation between the exponents $y$ and $\varepsilon$. The expressions for the critical exponents of the tracer field $\theta$ were obtained in the one-loop approximation. 

In order to find the anomalous exponents of the structure functions, the composite fields~(\ref{Fnp}) were renormalized. The critical dimensions of them were evaluated. It turned out that there is an infinite number of the dangerous operators, i.e., the operators with negative critical dimensions. Besides, OPE allowed us to derive the explicit expressions for the critical dimensions of the structure functions. 
The existence of the anomalous scaling in the inertial-convective range was established for both possible scaling regimes.
Another very interesting result is that some kinds of operators exhibit the ``multifractal'' behavior.

With regard to future research, it would be interesting to go beyond the one-loop approximation and to analyze the behaviour more precisely on the higher level of accuracy. To have a closer look at the vector fields,both passive and active, is also a challenge to be further investigated.

\begin{acknowledgement}

The authors are indebted to M.~Yu.~Nalimov, L.~Ts.~Adzhemyan, M.~Hnati\v{c}, J.~Honkonen, and V.~\v{S}kult\'ety for discussions.

The work was supported by VEGA grant No.~1/0345/17 of the Ministry
of Education, Science, Research and Sport of the Slovak Republic
and by the Russian Foundation for Basic
Research within the Project 16-32-00086.
N.~M.~G. acknowledges the support from the Saint Petersburg Committee of Science and High School.
N.~M.~G. and M.~M.~K. were also supported by the Dmitry Zimin's ``Dynasty'' foundation.

\end{acknowledgement}

\end{document}